\documentclass[twocolumn,showpacs]{revtex4}
\usepackage[xdvi]{graphicx}%
\usepackage{dcolumn}
\usepackage{amsmath}

\makeatletter
\def\btt#1{\texttt{\@backslashchar#1}}%
\DeclareRobustCommand\bblash{\btt{\@backslashchar}}%
\makeatother
\newcommand{\pder}[2]{\frac{\partial #1}{\partial #2}}
\newcommand{\pdert}[2]{\frac{\partial^2 #1}{\partial #2^2}}
\newcommand{\pderf}[2]{\frac{\partial^4 #1}{\partial #2^4}}
\newcommand{\bra}{\left\langle}
\newcommand{\ket}{\right\rangle}
\newcommand{\ksh}{h_{\rm ks}}

\begin{document}


\title[A fluctuation theorem for phase turbulence]
{A fluctuation theorem in phase turbulence 
of chemical oscillatory waves}

\title{A fluctuation theorem for phase turbulence 
of chemical oscillatory waves}

\author{Shin-ichi Sasa}%
\email{sasa@jiro.c.u-tokyo.ac.jp}
\affiliation{%
Department of Pure and Applied Sciences, 
University of Tokyo, \\
Komaba, Meguro-ku, Tokyo 153, Japan
}%

\date{\today}%

\begin{abstract}
Through numerical simulations of the Kuramoto equation, which displays 
high-dimensional dissipative chaos, we find a quantity representing 
the cost for maintenance of a spatially non-uniform structure that 
appears in the phase turbulence of chemical oscillatory waves. We call this
quantity the generalized entropy production and demonstrate that its 
distribution function possesses a symmetry expressed by a fluctuation 
theorem. We also report a numerical result which suggests a relation between 
this generalized entropy production rate  and the Kolmogorov-Sinai entropy.
\end{abstract}


\maketitle


%
%

When a system exists in a non-equilibrium steady state, the entropy produced 
in the system continuously flows into the environment.  The amount of the 
entropy produced during a finite time interval is called  the 'entropy 
production'. As stipulated by the second law of thermodynamics, the 
entropy production must be positive. 

Entropy production can be interpreted as the cost for the maintenance of 
non-equilibrium steady states.  This leads us to  expect that
we can find a quantity which can be regarded as a {\it generalized
entropy production} characterizing non-thermodynamic as well as 
thermodynamic systems. With the hope of demonstrating this point,
we study the phase turbulence of chemical oscillatory waves \cite{Kbook}.

When a pacemaker is situated in a small region in a spatially uniform  phase 
turbulent state, a spatially non-uniform state appears, as is easily
understood. In this paper, through numerical experiments, we find 
an expression for a generalized entropy production characterizing 
such a spatially non-uniform state.  We further attempt to relate this 
generalized entropy production with the Kolmogorov-Sinai (KS) entropy,
which measures the rate of information loss in chaotic dynamical systems.

%
%

We first introduce a mathematical model describing phase turbulence.
For simplicity, we consider reaction diffusion systems in a one-dimensional 
circuit. Such systems are considered as describing the behavior of the
concentrations of chemical species.  For a system of this type,
when  certain conditions are satisfied (see \cite{Kbook} for details),
their  concentrations  oscillate at each position $x$, and the 
phase $\phi$ of the oscillation varies slowly in time $t$. In the weakly
unstable case, the time evolution of the phase $\phi(x,t)$ is described by 
the Kuramoto equation \cite{Kbook,K-T},
\begin{equation}
\pder{\phi}{t}+\pdert{\phi}{x}+\pderf{\phi}{x}
 +\frac{1}{ 2} \left(\pder{\phi}{x}\right)^2=\epsilon \omega(x),
\label{KS}
\end{equation}
where we have carried out appropriate scalings for $x$, $t$ and $\phi$
so as to obtain the simplest form of the equation. The Kuramoto equation
(\ref{KS}) is defined over the region  $0 \le x \le L$, with periodic 
boundary conditions. 
The right-hand side of (\ref{KS}) represents the effect of external 
oscillation introduced by the pacemaker: $\epsilon$ corresponds to 
the pacemaker strength, and the form of $\omega(x)$ reflects its
position.  We assume that  $\omega(x)=1$ in the region 
$ 0 \le x \le H$ and $\omega(x)=0$ otherwise.  In the analysis below,
we study the case  $(L,H)=(128,10)$. 

%
%

We obtain an expression for the generalized entropy production by 
undertaking a numerical analysis of (\ref{KS}).
In our numerical simulations, we discretized space with a unit size 
$\delta x=1.0$ and employed an explicit Euler method  with  time 
step $\delta t=0.01$. The nonlinear term  
was discretized as $((p_{i+1}-p_{i-1})/2\delta x)^2/2$, where 
$p_i=\phi(i \delta x,t)$. 
We believe  that this numerical scheme is sufficient to allow
for investigation of the behavior of interest.

%
%

When there is no pacemaker ($\epsilon=0$), (\ref{KS}) possesses 
spatio-temporal chaotic solutions for a rather wide class of initial 
conditions \cite{Kbook,Y-K}. (Precisely speaking, there is a very small 
 basin for stable spatially periodic solutions \cite{Mann,Fri,Shr}. 
However, we do not  consider such solutions, because this basin
is too small to be observed when initial conditions are assigned 
randomly \cite{Shr}.)  The statistical properties of this spatio-temporal 
chaos have been extensively studied both numerically \cite{Y-K,Mann,stat} and
theoretically \cite{F-Y,Yakhot}.

\begin{figure}
\begin{center}
\includegraphics[width=8cm]{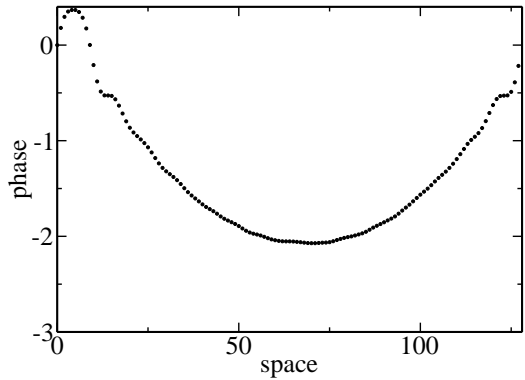}
\end{center}
\caption{
Long time average of the phase profiles, each of which is shifted 
so as to satisfy $\phi(0,t)=0$. Here, $\epsilon=0.1$. The averaging time 
was $4\times 10^7$ time units for the statistical steady state. 
}
\label{fig1}
\end{figure}

When $\epsilon \not= 0$, the system possesses spatially non-uniform 
statistical properties. As an example, in Fig. \ref{fig1}, we plot
the long-time average of the phase profiles, each of which is shifted 
so as to satisfy $\phi(0,t)=0$. 

%
%

Now, let us consider the  generalized entropy production which represents 
the cost for maintaining the spatially non-uniform state.  
It is important to note here that the generalized entropy production 
will not be expressed in terms of heat, in  contrast to the  usual entropy 
production
of non-equilibrium steady states, because the phase turbulent state 
cannot be described by quantities characterizing thermodynamic 
equilibrium states. For this reason, we  have no intention to 
base our analysis of the generalized entropy production on 
thermodynamic considerations. Instead, we consider a more abstract
condition that we expect the generalized entropy production to satisfy.

%
%

As a recent development in the study of non-equilibrium steady 
states \cite{Evans,G-C,Kur,Maes}, an interesting relationship 
regarding the entropy production has been established.
Defining $\Pi(z;\tau)$ to be the probability density 
that the entropy production $\Sigma_\tau$ during a finite time 
interval $\tau$ takes the value $z$ in a given  non-equilibrium 
steady state, it has been shown that this probability density
possesses the  symmetry
\begin{equation}
\log \Pi(z;\tau)-\log\Pi(-z;\tau) = z+o(\tau)
\label{fluc}
\end{equation}
in the limit  $\tau \rightarrow \infty$. This symmetry is referred 
to as  a 'fluctuation theorem'.  {}From (\ref{fluc}), it is easily 
derived  that the most probable value of the entropy production ratio 
$\Sigma_\tau/\tau$  in the large $\tau$ limit is positive. 
Considering the above result for entropy production in non-equilibrium 
steady states, we conjecture that a similar result can be obtained 
for a wider class of nonequilibrium states by appropriately defining
a generalized entropy production and seeking a generalized fluctuation
theorem it satisfies. 

%
%

The problem we face is to find a generalized entropy production 
characterizing  spatially non-uniform phase turbulent states, whose 
distribution function satisfies (\ref{fluc}).  By recalling that 
the entropy production in  a  non-equilibrium steady state can be 
expressed in the form [force][flux]/[temperature], and by interpreting 
$\epsilon \partial_x \omega$ and $\partial_x{\phi}$  as corresponding to a 
force and a flux, we regard the following quantity as a candidate 
of the generalized entropy production:
\begin{equation}
\Sigma_\tau^{\rm K}
=\epsilon \beta\int_0^\tau  dt \int_0^L  dx (-\omega(x))\pdert{\phi}{x},
\label{ep}
\end{equation}
where we note that the integrand in (\ref{ep}) is rewritten as 
$\partial_x \omega \partial_x{\phi}$. $\beta$ corresponds to the 
effective inverse temperature of the phase turbulence whose value 
is determined later so that  the distribution function of 
$\Sigma_\tau^{\rm K}$  satisfies (\ref{fluc}).

\begin{figure}
\begin{center}
\includegraphics[width=8cm]{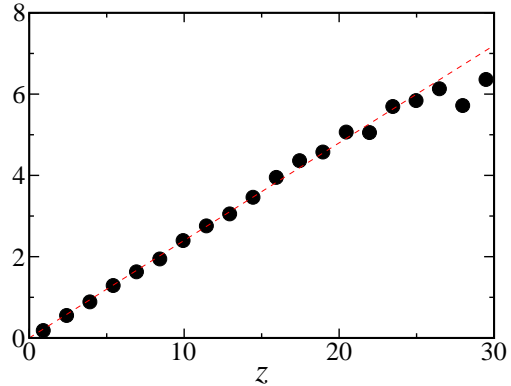}
\end{center}
\caption{
$\log \Pi(z;\tau,1)-\log \Pi(-z;\tau,1)$.
The dotted line corresponds to $\log \Pi(z;\tau,1)-\log \Pi(-z;\tau,1)
=0.24z.$ Here, $\epsilon=0.1$ and $\tau=200$. The distribution was 
made by using $2\times 10^5$ trajectory segments for the statistical 
steady state. 
}
\label{fig2}
\end{figure}
 
Through numerical experiments, we obtained the distribution function 
$\Pi(z;\tau,\beta)$ of $\Sigma_\tau^{\rm K}$ for a statistically steady 
state. Since the value of $\beta$ is not determined yet, we first 
considered an arbitrary value and calculated $\Pi(z;\tau,1)$. As seen
from Fig. \ref{fig2}, it seems that the equality
\begin{equation}
\log \Pi(z;\tau,1)-\log \Pi(-z;\tau,1) = 0.24z
\label{fluc1}
\end{equation}
holds approximately. Then, since the trivial identity
\begin{equation}
\Pi(z;\tau,\beta)=\Pi(z/\beta;\tau,1)/\beta
\end{equation}
holds, $\Pi(z;\tau,\beta)$  with $\beta=0.24$  satisfies 
the fluctuation theorem (\ref{fluc}).

\begin{figure}
\begin{center}
\includegraphics[width=8cm]{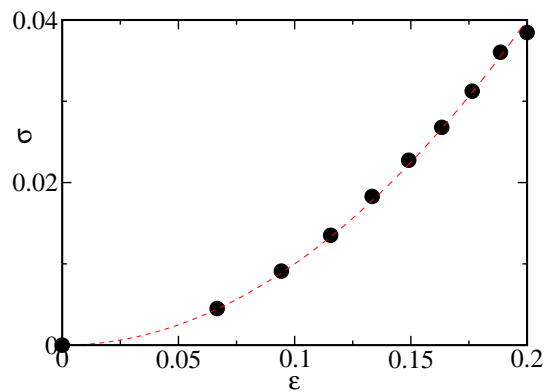}
\end{center}
\caption{
Entropy production rate $\sigma$ as a function of $\epsilon$. 
The dotted line corresponds to $\sigma=\epsilon^2$.
The averaging 
time was $4 \times 10^5$ time units for the statistical steady states.
}
\label{fig3}
\end{figure}

If we wish to characterize  spatially non-uniform phase 
turbulence by a single quantity, the generalized entropy production ratio,
\begin{equation}
\sigma =\lim_{\tau \rightarrow \infty}  \frac{\Sigma_\tau^{\rm K}}{\tau},
\end{equation}
may be most important. 
In Fig. \ref{fig3}, we plot $\sigma$ as a function of
the pacemaker strength $\epsilon$.  As is true generally for
non-equilibrium steady states, we see that $\sigma$ is proportional to 
$\epsilon^2$ in the region of small $\epsilon$.

%
%

We can formulate an  argument to support our numerical result by
considering the Yakhot conjecture \cite{Yakhot,Zaleski,Sne,Chow} 
that the long time and large distance behavior of statistical properties 
of chaotic solutions to (\ref{KS}) are equivalent 
to that of solutions to the stochastic evolution equation (which is 
essentially the same as the Burgurs equation with noise)
\begin{equation}
\pder{\tilde \phi}{t}-\nu\pdert{\tilde \phi}{x}
 +\frac{\lambda}{2} \left(\pder{\tilde \phi}{x}\right)^2=\epsilon\omega(x)+\xi,
\label{nbur}
\end{equation}
where $\xi$ is a Gaussian white noise satisfying 
\begin{equation}
\bra \xi(x,t)\xi(x',t') \ket=2D \delta(t-t')\delta(x-x').
\label{np}
\end{equation}
Here, $\nu$ and $D$ are positive constants. It is quite easy to demonstrate
a fluctuation theorem for this stochastic
system (See Refs. \cite{Kur,Maes}).  The result is the following: 
When we define the entropy production as 
\begin{equation}
\Sigma_\tau^{\rm B}=\epsilon\frac{\nu}{D}
\int_0^\tau  dt \int_0^L  dx (-\omega(x))\pdert{\tilde \phi}{x},
\label{ep2}
\end{equation}
the distribution function of $\Sigma_\tau^{\rm B}$ 
for a statistical steady state satisfies the fluctuation
theorem  (\ref{fluc}).  Therefore, if the Yakhot conjecture is valid
and if this conjecture ensures the equality of the space-time inetegration 
in (\ref{ep}) with that in (\ref{ep2}), the expression of (\ref{ep}) 
should hold with  $\beta=\nu/D$. 

%
%

We would like to consider the generalized entropy production from 
the  dynamical system viewpoint.  Until now, 
we have not succeeded in deriving the fluctuation theorem 
in a manner based on dynamical system considerations.  
However, we have obtained  a numerical  result concerning 
the relation between the generalized entropy production rate and 
the KS entropy, which can be calculated using the Lyapunov analysis.  
We now discuss this result.
\begin{figure}
\begin{center}
\includegraphics[width=8cm]{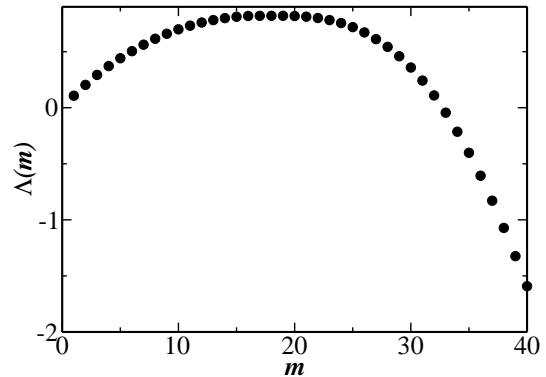}
\end{center}
\caption{
Long time average of the  expansion ratio of  $m$-dimensional volume elements.
The averaging time was $10^4$ time units for the statistical steady state.
}
\label{fig4}
\end{figure}

\begin{figure}
\begin{center}
\includegraphics[width=8cm]{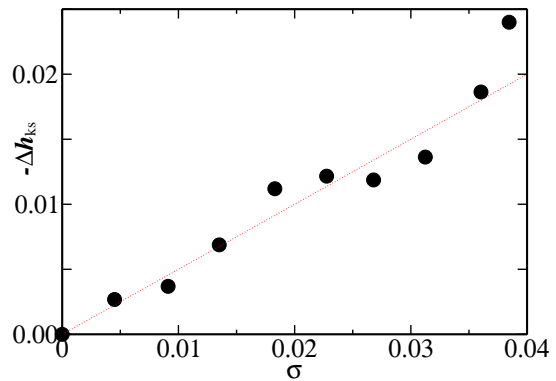}
\end{center}
\caption{
$(\sigma, -\Delta \ksh)$ for different values of $\epsilon$
$(0 \le \epsilon \le 0.2)$. The dotted line corresponds to
$-\Delta \ksh=\sigma/2.$ The averaging time 
was $4 \times 10^5$ time units for the statistical steady states.
}
\label{fig5}
\end{figure}

Performing  numerical integrations of the linearized equation 
corresponding to (\ref{KS}) for  chaotic solutions and employing 
the Gram-Schmidt decomposition technique \cite{SN}, we calculated 
the long time average of the 
expansion ratio, $\Lambda(m)$, for  $m$-dimensional volume 
elements along chaotic trajectories \cite{Mann}.  
In Fig. 4, the graph of $\Lambda(m)$  is plotted  for the system without 
the pacemaker.  The KS entropy is defined as the maximal value of 
$\Lambda(m)$.
Let $\ksh(\epsilon)$ be the KS entropy for the system with  pacemaker 
strength $\epsilon$. Since we  naively expect that $\sigma(\epsilon)$
is related to the quantity 
\begin{equation}
\Delta \ksh(\epsilon)=\ksh(\epsilon)-\ksh(0),
\end{equation}
we plot  points $(\sigma, -\Delta \ksh)$ for different values of 
$\epsilon$. As displayed in Fig. \ref{fig5}, the  results indicate
the relation 
\begin{equation}
-\Delta \ksh=\frac{1}{2} \sigma
\label{rel}
\end{equation}
for small $\epsilon$. Note that we observed a greater deviation 
from (\ref{rel}) when $\epsilon$ is increased.

We do not  understand  why the numerical factor ${1/2}$ appears 
in (\ref{rel}).  However, the fact we have obtained such a simple
relationship for the system we consider presently may suggest 
the existence of a universal relation between the generalized 
entropy production and some dynamical system quantity. (In the study of a 
time-dependent Hamiltonian system, the same factor appears in the
relation  between the Boltzmann entropy difference and the excess information 
loss found through Lyapunov analysis \cite{SK1}.)
Further investigation is necessary to clarify this point.


We now make some remarks.

A fluctuation theorem  was first demonstrated numerically \cite{Evans}
and proved  mathematically \cite{G-C} in the study of a deterministic 
Gaussian thermostatted model.  In this model,  
the entropy production is given by the phase space contraction, and 
the fluctuation theorem is derived by exploiting the time-reversibility 
of the deterministic  evolution equation. Indeed in many cases,
including that of the stochastic model (\ref{nbur}) with (\ref{np}), 
the entropy production for a given trajectory has been related to 
the ratio of the measure of this trajectory to that of its time 
reversal trajectory.
However, such an argument cannot be applied to (\ref{KS}), 
because it does not possess time-reversal symmetry. 

On a more abstract level, as discussed by Maes \cite{Maes}, 
fluctuation theorems can be associated with a transformation $P$,
that  acts on trajectories and satisfies $P^2=1$. 
In certain situations (e.g. for steady states), this transformation
is equivalent to time reversal, but in general this is not the case.
The problem is thus to determine the form that $P$ takes in the
present case and to determine how to derive our fluctuation theorem
from it. 

Finally, we briefly discuss the size dependence for the system we have 
studied. It is known that the statistical properties of phase turbulence 
without a pacemaker depend on $L$ in an anomalous way 
\cite{F-Y,Yakhot,Zaleski,Sne,Chow}. The system size in our numerical  
experiments is much smaller than the size representing cross-over
to the region in which the anomalous behavior is observed. It is
an interesting problem to extend our study to this anomalous scaling region.

%
%

In summary, we have demonstrated that the generalized entropy production 
(\ref{ep}), which has been obtained phenomenologically in the consideration
of spatially non-uniform phase turbulent states, satisfies the fluctuation 
theorem (\ref{fluc}).  Unfortunately, there is no obvious way to derive 
this fluctuation theorem from the deterministic evolution equation in the 
present case,  because we cannot employ an argument relying on
time-reversibility. Nevertheless, using the Yakhot conjecture,
we have formulated a reasonable explanation of the result.  We have also 
obtained the  numerical result (\ref{rel}), which suggests an interesting 
relation between the generalized entropy production ratio and the KS 
entropy.  We believe that a deeper understanding of the problem 
considered here will allow for the study of dissipative high-dimensional 
systems, such as fluid turbulence and granular flow, from a new point
of view. 

%
%

The author acknowledges Y. Kuramoto for explaining  the phenomenon 
considered  presently fourteen years ago. He also thanks H. Kamai 
for  collaboration on a preliminary part of the study, and G. C. Paquette 
for critical reading of the manuscript and for his comments.
This work was supported by grants from the Ministry of Education, 
Science, Sports and Culture of Japan, Nos. 12834005 and  11CE2006.

%
%


\end{document}